\documentclass
[prb,twocolumn,showpacs]{revtex4}
\usepackage{graphicx}
\usepackage{amsfonts}
\usepackage{amssymb}

\newcommand{\ba}{\begin{eqnarray}}
\newcommand{\be}{\begin{equation}}
\newcommand{\ea}{\end{eqnarray}}
\newcommand{\ee}{\end{equation}}

\newcommand{\ignore}[1]{}

\begin{document}

\title{How does chemical
functionalization affect the thermal transport of antimony films?}

\author{Tian Zhang$^{1}$}

\author{Yuan-Yuan Qi$^{2}$}

\author{Xiang-Rong Chen$^{1}$}
\email{xrchen@scu.edu.cn}

\author{Ling-Cang Cai$^{3}$}

\address{$^{1}$Institute of Atomic and Molecular Physics, College of Physical Science and Technology, Sichuan
             University, Chengdu 610065, China\\$^{2}$College of Science, Henan University of Technology, Zhengzhou 450001, China\\$^{3}$National Key Laboratory for Shock Wave and Detonation Physics Research, Institute of Fluid Physics, Chinese Academy of Engineering Physics, Mianyang 621900, China }

%\author{Tian Zhang$^{1}$, Zhao-Yi Zeng$^{2}$,Xiang-Rong Chen$^{1}$, and Ling-Cang Cai$^{3}$}

%\email{ycheng@scu.edu.cn; xrchen@scu.edu.cn}

%\affiliation{$^{1}$Institute of Atomic and Molecular Physics, College of Physical Science and Technology, Key
%Laboratory of High Energy Density Physics and Technology of Ministry of Education, Sichuan
%University, Chengdu 610065, China}

%\affiliation{$^{2}$College of Physics and Electronic Engineering, Chongqing Normal University, Chongqing 400047, China}

%\affiliation{$^{3}$National Key Laboratory for Shock Wave and Detonation Physics Research, Institute of Fluid Physics, %Chinese Academy of Engineering Physics, Mianyang 621900, China}

\date{June 26, 2016}

\begin{abstract}
Chemical functionalization is an effective means to tune electronic and crystal structure of two-dimensional material, but very little is known regarding its correlation with thermal transport. Based on the first-principle calculation and an iterative solution of Boltzmann transport equation, we find that antimony films are potential excellent thermoelectrical materials with rather low thermal conductivities $k$ ($<$ 2.5 W/mK), and chemical functionalization can induce the reduction in $k$ to some extent, which is mainly due to the reduction of phonon lifetimes limited by the anharmonic scattering. More interesting, the origin of the reduction in $k$ is not the anharmonic interaction but the harmonic interaction from the depressed phonon spectrum mechanism, and for some chemical functional atom in halogen, flat modes appearing in the low frequency range play also a key factor in the reduction of $k$ by significantly increasing the three-phonon scattering channels. Our work provides a new view to adjust the thermal transport which can benefit thermoelectric material design, and analyzes the reduction mechanism in $k$ from the chemical functionalization for antimonene.
\end{abstract}


\maketitle
Two-dimensional (2D) materials have been a focus of intense research\cite{TW1,TW2,TW4} since the discovery of graphene with high electron mobility\cite{AJ1}, and very high thermal conductivity\cite{AJ2,AJ3}. More 2D materials, such as  silicene, phosphorene, and transition metal dichalcogenides, are now considered for various practical usages due to their distinguished properties. Recently, a new and promising 2D semiconductor, antimonene in the same group of the periodic table with phosphorene, was proposed on the basis of first-principles calculations\cite{an1}. Its isolation was expected to be achieved by exfoliating layered Sb crystal, which is due to the weak interlayer interaction. Similar to silicene, antimonene has also a non-planar honeycomb-like structure, and its electronic structure can also show the topological character under external strain\cite{to1}. The character of topological insulator for antimonene is useful to thermoelectrical application with perfectly conducting channels for electrons, thanks to its non-dissipative edge states\cite{td2}. A high thermoelectric performance requires the system to be a bad conductor for phonon but a good conductor for electron. Compared to other widely studied 2D materials, the thermal transport properties of monolayer antimonene is still not well-understood.

Generally both electrons and phonons contribute to the total thermal conductivity $k$ in nonmagnetic crystal\cite{ep1}, with phonons dominating in semiconductor and insulators. Electron transport can be optimized by adjusting the geometry size to maximize the contribution of non-dissipative edge states for 2D topological insulators, and phonon transport is usually be tuned by external pressure\cite{pre2,pre3,pre4}. The theory of Liebfried and Schl\"{o}mann(LS)\cite{LS} have typically been approved for the pressure dependence of $k$ in nonmetallic crystalline materials, which predicts that $k$ must increase with $P$\cite{kp}, i.e., $dk/dP$ $>$ $0$. Filler is also excellent method to adjust phonon transport for 3D materials\cite{st,st1}, and some relative theories have also been raised, such as a rattling mechanism postulated by Slack and Tsoukala to explain the reduction of $k$ in filled skutterudites\cite{st}, which implies the fillers move independently from the host matrix (Einstein motion) and effectively scatter phonons. Like to filler, the chemical functionalization is also a universal way to tune mechanical, electronic and other properties for 2D materials. Furthermore, the chemical functionalization can even induce the topological phase transition, which is found by Xu $et$ $al$\cite{XU}. However, the influence for the chemical functionalization and corresponding mechanism in thermal transport properties are still uninvestigated.

\begin{figure*}
\includegraphics[width=6.2in]{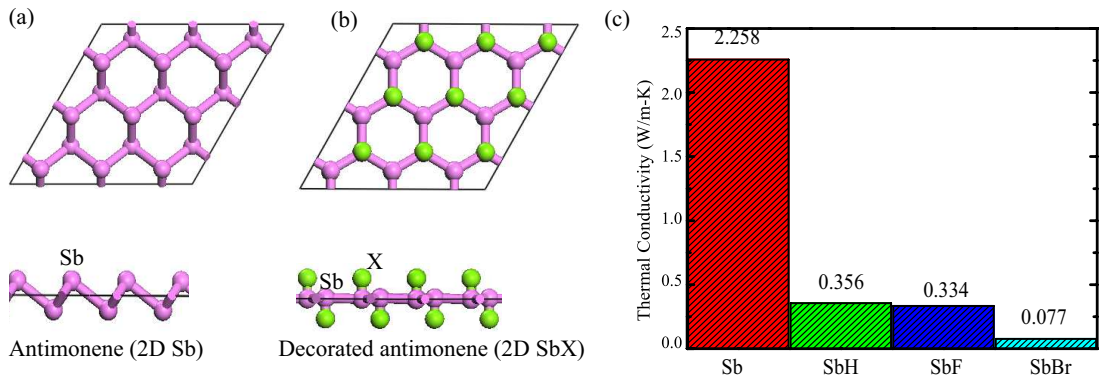}
\caption{(Color online) Top and side view of (a) the antimonene structure and (b) the decorated antimonene structure. $X$ represents the chemical functional group. (c) The calculated lattice thermal conductivities $k$ for antimonene and decorated antimonene.}
\label{fig.sketch}
\end{figure*}

In this work, we calculate the lattice thermal conductivities $k$ of antimony films using the first-principle and an iterative solution of Boltzmann transport equation (BTE) for phonons\cite{BTE}. The rather low $k$ ($<$ 2.5 W/mK) are found in these antimony films, which is very favorable to realize high thermoelectrical efficiency. More interesting, the chemical functionalization can reduce the value of $k$ compared with the antimonene to some extent, and the $k$ becomes lower and lower along with the mass increase of chemical functional element . We make a detailed discussion for this difference in $k$, and find that the anharmonic third-order interatomic force constants play a trivial role. This difference can be due to the depressed phonon spectrum by analyzing the weighted phase space $W$, and the flat phonon modes is also a key factor in the reduction of $k$ by increasing more scattering channels when the chemical functional atom is in halogen.

The first-principle calculations based on the density functional theory are performed using the Vienna ab simulation package (VASP)\cite{vasp1,vasp2} . We use the generalized gradient approximations (GGA) of Perdew-Burke-Ernzerhof (PBE)\cite{PBE} for electron-electron interactions and the projector-augmented-wave (PAW)\cite{PAW} pseudo-potentials in the plane-wave basis with an energy cutoff of 700 eV. The Brillouin zone (BZ) was sampled using a 13$\times$13$\times$1 gamma-centered Monkhorst-Pack grid\cite{MK} during structural relaxation for the unit cell. A slab model, together with a vacuum layer larger than 30 {\AA}, was employed to avoid spurious interactions due to the nonlocal nature of the correlation energy\cite{slab}.

Thermal conductivity of Sb monolayer is calculated using BTE with relaxation time approximation as implemented in ShengBTE package\cite{BTE}, where thermal conductivity is given by\cite{termal}
\begin{equation}
    k_{\alpha}=1/N_{q}V\sum\limits_{q,j}C_{q,j}\upsilon^{2}_{q,j,\alpha}\tau_{q,j},
\end{equation}
where $N_{q}$ is the number of uniformly spaced q points in the Brillouin zone, $V$ is the volume of the unit cell, $C_{q,j}$, $\upsilon_{q,j,\alpha}$ and $\tau_{q,j}$ are special heat, group velocity along transport direction $\alpha$, and relaxation time of the phonon mode with wave vector q and polarization $j$, respectively. In the calculation of phonon dispersion, the harmonic second-order IFCs are obtained using density functional perturbation theory (DFPT) as implemented in the PHONOPY code\cite{FP} with 5$\times$5$\times$1 k-mesh. The ShengBTE package is used to obtain the anharmonic third-order IFCs, and solve the BTE which needs harmonic second- and anharmonic third-order interatomic force constants (IFCs) as inputs. The ShengBTE package is completely parameter-free and based only on the information of the chemical structure. Full solution of BTE is also obtained by using an iterative scheme, and it is found that the relaxation time approximation (RTA) has a very good accuracy.

The optimal lattice structures of antimonene monolayer and decorated antimonene monolayer (denoted as Sb and SbX) are shown in Fig. 1(a) and 1(b), respectively, where X represents the chemical functional group. The lattice constant is 4.139 {\AA} for Sb, slightly longer than other theoretical results\cite{an1,zhao1}, 5.279 {\AA} for SbH, 5.144 {\AA} for SbF and 5.245 {\AA} for SbBr. Compared to Sb, in SbX the Sb-Sb bond length slightly increases, and interestingly, different from the low-buckled configuration for the 2D Sb, the chemical functionalization of the buckled Sb induces its structure to transition into an extremely 2D sheet [see Fig. 1(b)], which is similar to the phenomenon in fully hydrogenated arsenene\cite{AS}. Fig. 1(c) shows the obtained lattice thermal conductivities $k$ for all antimony films based on the iterative solution at 300 K. The $k$ of these antimony films are universally low, which is very favorable to realize high thermoelectrical efficiency, especially as topological insulator under external strain\cite{to1}. The $k$ about 2.258 W/mK for Sb is small compared with the bulk Sb ($\sim$ 10-60 W/mK for 50-300 K\cite{BULK}). This is reasonable because phonons are more sensitive to boundary scattering when the thickness of the Sb thin film decreases, and thus further reduce $k$. In order to check the reliability of the $k$ for Sb, we change the supercell size from 3$\times$3$\times$1 to 5$\times$5$\times$1 to obtain more accurate anharmonic third-order IFCs, and simultaneously reexamine the harmonic second-order IFCs within density functional perturbation theory using the QUANTUM ESPRESSO package (QE)\cite{QE1,QE2}. Results indicate that $k$ is not affected highly by the supercell size with error lower than 0.1 W/mK, and the error of $k$ is about 0.4 W/mK between VASP and QE packages. We make also a comparison with other 2D hexagonal materials, and their $k$ are as follows: MoS$_{2}$ $\approx$ 4-8 W/mK\cite{MOS2}, blue phosphorene $\approx$ 78 W/mK\cite{bp}, silicene $\approx$ 7-11 W/mK\cite{si}, stanene $\approx$ 11.6 W/mK\cite{sn}and arsenene $\approx$ 7.8 W/mK (armchair direction)\cite{as}. Although the obtained $k$ for Sb is smaller than those for other same group (blue phosphorene and arsenene), this discrepancy can be accepted because the larger mass usually leads to the lower $k$ based on LS relation\cite{LS}.

\begin{figure}
\includegraphics[width=2.4in]{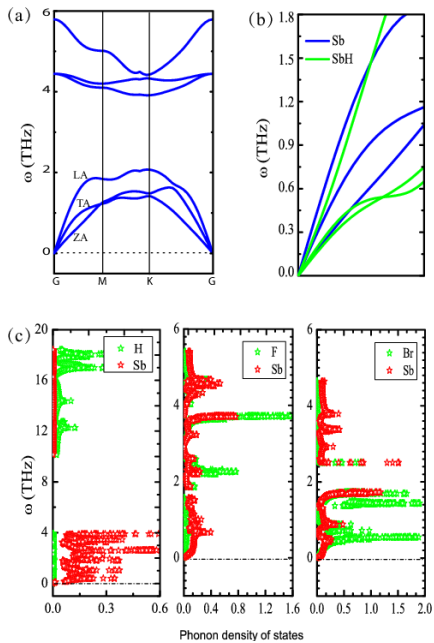}
\caption{(Color online) Phonon spectrum $\omega(\vec{k})$ of (a) the antimonene Sb. ZA and TA are the transverse acoustic phonon branches, and LA is the longitudinal acoustic phonon branch. G, M and K refer to special points in the first Brillouin zone. (b) The enlarged view of phonon spectrums for the antimonene Sb and decorated antimonene SbH. (c) The phonon density of states (PDOS) for SbH, SbF and SbBr.}
\label{fig.sketch}
\end{figure}

An interesting result is found in Fig. 1(c) that the $k$ values for all SbX films are lower than that for Sb, and the $k$ becomes lower and lower along with the mass increase of chemical functional element. Phonon spectrums and phonon density of states (PDOS) of the antimony films are presented in Fig. 2, where the phonon spectrum of the Sb is well consistent with that by Wang $et$ $al$\cite{wang}. All lattice structures are thermodynamically stable and their stability do not depend on the substrates. It is noted that the low frequency range has a less partial PDOS of H atoms, and the high frequency range is dominated by the vibrations of H atoms. However, along with the mass increase of chemical functional element, the partial PDOS of chemical functional atoms are more and more in the low frequency range, and them become even the dominant ingredients for SbBr, as shown in Fig. 2(c). Hence, next we will elucidate the role of the chemical functionalization in the reduction of $k$ by considering conventional anharmonic scattering along.

As we know, $C_{\nu}$, $\upsilon_{\lambda}$ and $\tau_{\lambda}$ can influence the value of $k$. For most of the heat conductive modes $C_{\nu}$ approaches the classic value $k_{B}$ at room temperature, and thus the $C_{\nu}$ can not influence $k$ value although the more number of phonon modes may lead to the total heat capacity increase slightly. We make a particular comparison for $\upsilon_{\lambda}$ between Sb and SbH in low frequency phonon dispersion range, as shown in Fig. 2(b). The calculated speed of sound along the G-M direction in the Sb, $\upsilon_{LA}$ =$7.5$ Km/s, is higher about 1.42 time than that value of 5.3 Km/s for SbH, and the values of $\upsilon_{\lambda}$ for remainder two acoustic branches are also higher slightly for Sb than those for SbH. The difference of $\upsilon_{\lambda}$ is so small to explain the reduction of $k$. Hence we believe that $\tau_{\lambda}$ should be the main factor for the decrease of $k$ in all SbX, and thus make a detailed discussion about their anharmonic scattering.

\begin{figure*}
\includegraphics[width=6.0in]{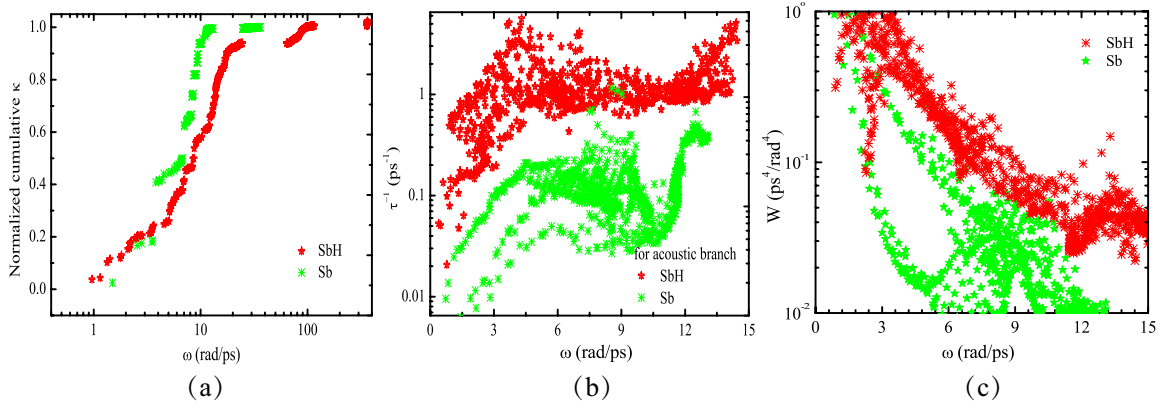}
\caption{(Color online) (a) The normalized cumulative $k$ for the Sb and SbH monolayers. (b) Anharmonic scattering rates SRs of the Sb and SbH monolayers in the frequency range of three acoustic branches. (c) Weighted phase space $W$, defined in Eq. (4) and (5), vs angular frequency for the Sb and SbH monolayers.}
\label{fig.sketch}
\end{figure*}

We neglect the isotope scattering and only take into account the anharmonic scattering for simplicity sake. The total scattering rate is equal to the inversion of the phonon lifetime $\tau_{\lambda}$, which can be obtained as the sum of three-phonon transition probabilities $\Gamma^{\pm}_{\lambda\lambda^{'}\lambda^{''}}$.  $\Gamma^{\pm}_{\lambda\lambda^{'}\lambda^{''}}$ can be expressed as\cite{t1,t2,t3}

\begin{equation}
    \Gamma^{\pm}_{\lambda\lambda^{'}\lambda^{''}}=\left(  \begin{array}{c}
    f_{\lambda^{'}}-f_{\lambda^{''}}\\
    f_{\lambda^{'}}+f_{\lambda^{''}}+1\\
                                                         \end{array}
    \right)
     \frac{h\pi\delta(\omega_{\lambda}+\omega_{\lambda^{'}}-\omega_{\lambda^{''}})}{4\omega_{\lambda}\omega_{\lambda^{'}}\omega_{\lambda^{''}}}|V^{\pm}_{\lambda\lambda^{'}\lambda^{''}}|^{2},
\end{equation}
where $\Gamma^{+}_{\lambda\lambda^{'}\lambda^{''}}$ corresponds to absorption processes, $\Gamma^{-}_{\lambda\lambda^{'}\lambda^{''}}$ describes emission processes, and $\delta$ function represents the energy conservation. $V^{\pm}_{\lambda\lambda^{'}\lambda^{''}}$ is the scattering matrix elements, which depends the anhamonic third-order IFCs $\phi^{\alpha\beta\gamma}_{ijk}$ \cite{t3}as

\begin{equation}
   V^{\pm}_{\lambda\lambda^{'}\lambda^{''}}=\sum\limits_{i\in u.c.}\sum\limits_{j,k}\sum\limits_{\alpha\beta\gamma}\phi^{\alpha\beta\gamma}_{ijk}\frac{e^{\alpha}_{\lambda}(i)e^{\beta}_{p^{'},\pm
   q^{'}}(j)e^{\gamma}_{p^{''},-q^{''}}(k)}{\sqrt{M_{i}M_{j}M_{k}}},
\end{equation}
where $M_{i}$ is the mass of atom $i$, and $e^{\alpha}_{\lambda}(i)$ is the $\alpha$ component of the eigenfunction of mode $\lambda$ at the $i$th atom. The contributions from different phonon branches to $k$ for Sb and SbH are analyzed. Like to the case in graphene, the total contribution of acoustic branches is more than 99$\%$ for Sb, implying that the contribution of optic phonons to $k$ is negligible. The contribution of LA is more than 57$\%$ and the contribution of ZA is less than 13$\%$, which is completely opposite compared with graphene (8$\%$ LA and 80$\%$ ZA) but is very similar to that in stanene\cite{sn}. This especial phenomenon in graphene is due to a symmetry selection rule arising from the reflection symmetry perpendicular to the graphene plane, which strongly restricts the participation of odd number of ZA phonons in three-phonon processes, e.g. ZA+ZA$\longleftrightarrow$ZA, ZA+LA/TA$\longleftrightarrow$LA/TA. However, the reflection symmetry in Sb is broken, which means that the selection rule does not apply. It is also not suited in SbH, and thus the contribution of TA to the $k$ is still little ($<$ 17$\%$). The normalized cumulative $k$ for the Sb and SbH, which is the fraction of heat conducted by the phonons with frequencies below a given value, is shown in Fig. 3(a). The normalized cumulative $k$ in the two cases are quite similar in low frequency range below 10 rad/ps. These phonons below 10 rad/ps contribute almost whole of the $k$ for Sb, and the phonons below 20 rad/ps contribute the most of $k$. Thus, next we need only to analyze phonon scattering rates (SRs) in low frequency range to look into the effect of $\tau_{\lambda}$ to $k$.

The SRs, $\tau_{\lambda}^{-1}$, is plotted in Fig. 3(b) versus $\omega_{\lambda}$ for Sb and SbH. It is obvious that the SRs for SbH is nearly larger than that for Sb in whole low frequency range, roughly by a factor of $\sim$ 6.0 in all range (not shown). This confirms that the reduction of $\tau_{\lambda}$ is the main reason for the reduced $k$ in SbH. The overall reduction of $\tau_{\lambda}$ and the main PDOS of H atom beyond 60 rad/ps in SbH overturns the resonant scattering theory\cite{rest}, which indicates that $\tau$ is reduced in a narrow frequency range i.e. the frequency of the guest atom vibrational mode. Anharmonic scattering arises mainly from the anharmonic interaction between atoms, and thus the additional anharmonic interaction introduced by the guest atom (H atom) may be the main reason for the reduction of $\tau_{\lambda}$. Here the H-related anharmonic interaction will not affect low frequency phonons below 15 rad/ps, because these vibrations do not involve H atoms observed by PDOS in Fig. 2(c). In additional, the harmonic second-order IFCs of Sb and SbH have not an excellent transferability and we do not interchange the anharmonic third-order IFCs to investigate whether the change of SRs comes from the anharmonic third-order IFCs. Hence, we can only research the contribution of the harmonic interaction described by harmonic second-order IFCs to the reduction in SRs. Weighted phase space $W$ provide by Li $et$ $al$\cite{st1} is the sum of frequency-containing factors in the expression of transition probabilities, where $V^{\pm}_{\lambda\lambda^{'}\lambda^{''}}$ is assumed unchange in decorated antimonene films, written as\cite{st1}

\begin{equation}
    W^{\pm}=\sum\limits_{\lambda^{'}p^{''}}\left(  \begin{array}{c}
    f_{\lambda^{'}}-f_{\lambda^{''}}\\
    f_{\lambda^{'}}+f_{\lambda^{''}}+1\\
                                                         \end{array}
    \right)\frac{\delta(\omega_{\lambda}+\omega_{\lambda^{'}}-\omega_{\lambda^{''}})}{\omega_{\lambda}\omega_{\lambda^{'}}\omega_{\lambda^{''}}},
\end{equation}

\begin{equation}
    W=W^{+}+W^{-}.
\end{equation}

$W$ is highly sensitive to the change of phonon spectrum, i.e. harmonic second-order IFCs, and the harmonic second-order IFCs can affect indirectly the SRs by the frequencies and the eigenvectors. Hence, $W$ is an excellent indicator to assessment the contribution to SRs from harmonic interaction. The calculated $W$ for Sb and SbH is plotted in Fig. 3(c). Almost for the whole low frequency range below 15 rad/ps the $W$ is apparently higher for SbH than that in Sb. Moreover, the difference in $W$ is well similar the difference in SRs for SbH with a similar factor of $\sim$ 5.9 on average. The agreement of the difference in $W$ and in $\tau^{-1}$ suggests that the change in frequency indeed accounts for the decrease of $\tau$. When the external temperature is much higher than the Debye temperature $\Theta_{D}$ ($\Theta_{D}$ for all SbX is $<$ 193 K), $f_{\lambda}$ in Eq. (4) can be approximated by $\frac{1}{exp(\hbar\omega_{\lambda}/k_{B}T)-1}$ $\approx$ $k_{B}T/\hbar\omega_{\lambda}$, and thus the $W$ is $\sim$ $\omega_{\lambda}^{-4}$. If the whole phonon spectrum is scaled by a factor of $c$, then $W$ for all modes are scaled by a factor of $c^{-5}$. We choose the reduction proportion of low phonon frequency nearby gamma point as the value of $c$, i.e. the ratio of $\upsilon_{LA}$, which is reliable because the set value of $c$ $\sim$ 1.13 in other work is very close to the ratio of group velocity $\sim$ 1.1\cite{st1}. From the previous discussion, the phonon spectrum of Sb is about 1.42 (c=1/1.42) times for that in SbH, and thus we have $c^{-5}$=5.77, which agrees with the difference of the $W$ observed in Fig. 3(c). This result makes us believe that the depressed phonon spectrum is the main origin of the increase of scattering in SbH.

\begin{figure}
\includegraphics[width=2.6in]{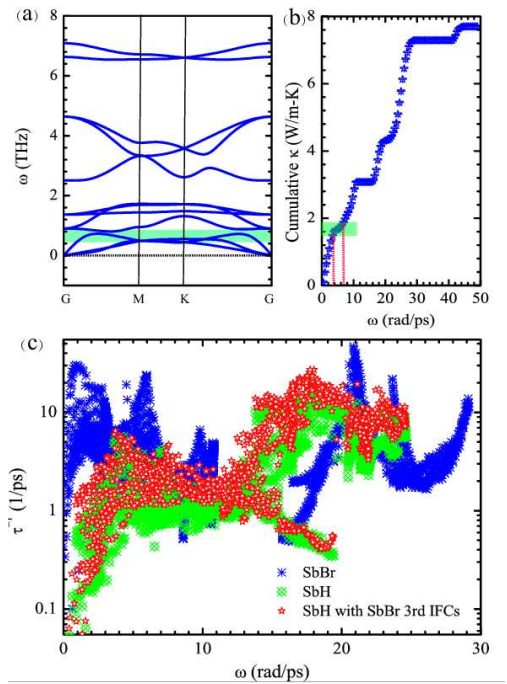}
\caption{(Color online) (a) The phonon spectrum and (b) the cumulative $k$ of the SbBr monolayer. The green shadowed area makes the flat modes. G, M and K refer to special points in the first Brillouin zone. The scale of ordinate in (b) is magnified 100 times. (c) Anharmonic scattering rates SRs of the SbBr and SbH monolayers compared with those of SbH monolayers calculated by using the anharmonic third IFCs of SbBr monolayer vs angular frequency.}
\label{fig.sketch}
\end{figure}

Based on the discussion in the PDOS of the antimony films [Fig. 2(c)], we know that the distribution of PDOS and the phonon spectrum for SbH are enormous different from those in SbX (X=F and Br). We may guess whether the effect of the chemical functional atom in halogen to the reduction in $k$ is different to that in the H atom, and thus make a specific comparison between SbH and SbBr. The phonon spectrum of SbBr is plotted in Fig. 4(a). Unlike to SbH, the SbBr displays noticeably flat modes in the frequency range from 0.2 to 0.6 THz. Furthermore, the cumulative $k$ for SbBr is also highly different from those in Sb and SbH, as shown in Fig. 4(b). The curve of SbBr shows a inflection point and begins to slower increase in the intermediate frequencies range from 1.0 to 7.0 rad/ps, and in those frequency range the SRs have the obvious enhancing, as shown in Fig. 4(c). The contributions of different phonon branches are also calculated and we can find that the contribution of optical branches between 20 to 30 rad/ps is unexpectedly large ($\sim$ 55$\%$), which exceeds the value of 21$\%$ in three acoustic branches. The especial phenomenon in the SbBr monolayer can be due to the unique flat modes, which is dominated by the vibrations of guest atoms.

Some works\cite{11, YB} indicated that this flat modes might cause an increase in phonon scattering, and used a conceptual diagram of phonon dispersion relations to prove it. When the frequency of the guest modes ($E_{guest}$) is larger than the gap frequency between the acoustic and upper-lying optical phonon modes ($\Delta$), the Umklapp scattering can be one of important processes in the suppression of thermal conductivity\cite{11}. The value of $\Delta$ here is much less than the value of $E_{guest}$, and the $E_{guest}$ about 0.4 THz is also a small value which can increase the number of phonons contributing to the process. Based on these unique conditions, the process of ¡°acoustic + optical $\rightarrow$ optical phonon¡± transition is easily found in a very wide area for SbBr, and thus the three-phonon channels are increased in acoustic phonons.  We also consider the contribution of the anharmonic third-order IFcs to the difference (SRs or $k$) between SbBr and SbH. The $k$ of SbBr with third-order IFCs of SbH is obtained about 0.103 W/mK, and the $k$ of SbH with third-order IFCs of SbBr is obtained about 0.275 W/mK, which shows that their anharmonic third-order IFCs are very similar, and can not be the reason for the very different SRs or $k$. The SRs of SbBr is larger than those of SbH roughly by a factor of $\sim$ 3.0 and the obtained $c^{-5}$ is about 1.93, which indicates that the depressed phonon spectrum mechanism also play a role in the low $k$ for SbBr and the remained difference is mainly from this flat modes. For the SbF, its phonon spectrum is very similar to that for SbBr, and its depressed ratio of phonon spectrum to that for SbH are obtained about -1.1 (negative represents lift). Because the special flat modes in the SbF can significantly increase the three-phonon scattering channels, and further reduce $k$ to seme extent compared to SbH. Hence, it is reasonable that the $k$ of SbF is comparative with that for SbH monolayer.

The reduction mechanism of the chemical functionalization on $k$ established above is similar to that from the filler in 3D materials\cite{st1,YB}. F (light element) and Br (heaven element) atoms are the two typical halogen elements, and their phonon spectrums of SbX (X in halogen) are also very similar. Thus we believe that this reduction mechanism of the chemical functionalization on $k$ is not special for the studied decorated antimonene here but a common feature.

In summary, the thermal conductivities $k$ of antimonene Sb and decorated antimonene SbX have been predicted based on $ab$ $initio$ Boltzmann transport equation method. The values of $k$ for these antimony films are low compared with other 2D hexagonal structures, which is very favorable to realize high thermoelectrical efficiency and be a promising candidate for next-generation thermoelectrical devices. We have found that the chemical functionalization can induce the reduction of $k$ for the antimony films, and proven that this change is mainly from the anharmonic scattering rates SRs, and comes minority from group velocity. The anharmonic third-order interatomic force constants is not the origin of different SRs but the depressed phonon spectrum by analyzing the weighted phase space $W$. In additional, the flat phonon modes is also a key factor in the reduction of $k$ by increasing more scattering channels when the chemical functional atom is in halogen. Our works prove that antimony films are excellent thermoelectric materials, and provide a new means to adjust the $k$ which can benefit materials design.

The work was supported by the NSAF Joint Fund Jointly set up by the National Natural Science Foundation of China and the Chinese Academy of Engineering Physics (Grant Nos. U1430117, U1230201).

\end{document}